\documentstyle[12pt]{article}
\newcommand{\nc}{\newcommand}
\nc{\renc}{\renewcommand}

\nc{\half}{{\textstyle{1\over2}}}
\nc{\etal}{\mbox{\it et al. }}
\nc{\ie}{{\it i.e.}}
\nc{\eg}{{\it e.g.}}

\renc{\thefootnote}{\arabic{footnote}}
\nc{\capt}[1]{{\bf Figure.} {\small\sl #1}}

\nc{\eqs}[2]{\mbox{Eqs.~(\ref{#1},\,\ref{#2})}}
\nc{\eq}[1]{\mbox{Eq.~(\ref{#1})}}

\nc{\figs}[2]{\mbox{Figs.~(\ref{#1},\,\ref{#2})}}
\nc{\fig}[1]{\mbox{Fig~.(\ref{#1})}}

\nc{\tag}[1]{\label{#1} \marginpar{{\footnotesize #1}}}
\nc{\mtag}[1]{\label{#1} \mbox{\marginpar{{\footnotesize #1}}}}
\renc{\baselinestretch}{1.5}
\newlength{\overeqskip}
\newlength{\undereqskip}
\nc{\be}[1]{\begin{equation} \mbox{$\label{#1}$}}
\nc{\bea}[1]{\begin{eqnarray} \mbox{$\label{#1}$}}
\nc{\Section}[2]{\section{#2}\label{#1}}
\nc{\Bibitem}[1]{\bibitem{#1}}
\nc{\Label}[1]{\label{#1}}

\nc{\eea}{\vspace{\undereqskip}\end{eqnarray}}
\nc{\ee}{\vspace{\undereqskip}\end{equation}}
\nc{\bdm}{\begin{displaymath}}
\nc{\edm}{\end{displaymath}}
\nc{\dpsty}{\displaystyle}
\nc{\bc}{\begin{center}}
\nc{\ec}{\end{center}}
\nc{\ba}{\begin{array}}
\nc{\ea}{\end{array}}
\nc{\bab}{\begin{abstract}}
\nc{\eab}{\end{abstract}}
\nc{\btab}{\begin{tabular}}
\nc{\etab}{\end{tabular}}
\nc{\bit}{\begin{itemize}}
\nc{\eit}{\end{itemize}}
\nc{\ben}{\begin{enumerate}}
\nc{\een}{\end{enumerate}}
\nc{\bfig}{\begin{figure}}
\nc{\efig}{\end{figure}}
\nc{\arreq}{&\!=\!&}
\nc{\arrmi}{&\!-\!&}
\nc{\arrpl}{&\!+\!&}
\nc{\arrap}{&\!\!\!\approx\!\!\!&}
\nc{\non}{\nonumber\\*}
\nc{\align}{\!\!\!\!\!\!\!\!&&}

\def\lsim{\; \raise0.3ex\hbox{$<$\kern-0.75em
      \raise-1.1ex\hbox{$\sim$}}\; }
\def\gsim{\; \raise0.3ex\hbox{$>$\kern-0.75em
      \raise-1.1ex\hbox{$\sim$}}\; }
\nc{\DOT}{\hspace{-0.08in}{\bf .}\hspace{0.1in}}
\nc{\Laada}{\hbox {$\sqcap$ \kern -1em $\sqcup$}}
\nc\loota{{\scriptstyle\sqcap\kern-0.55em\hbox{$\scriptstyle\sqcup$}}}
\nc\Loota{{\sqcap\kern-0.65em\hbox{$\sqcup$}}}
\nc\laada{\Loota}
\nc{\qed}{\hskip 3em \hbox{\BOX} \vskip 2ex}

\nc{\real}{{\rm I \! R}}
\nc{\Z}{{\sf Z \!\!\! Z}}
\nc{\complex}{{\rm C\!\!\! {\sf I}\,\,}}
\def\bigid{\leavevmode\hbox{\small1\kern-3.8pt\normalsize1}}
\def\id{\leavevmode\hbox{\small1\kern-3.3pt\normalsize1}}
\nc{\slask}{\!\!\!/}
\nc{\bis}{{\prime\prime}}
\nc{\pa}{\partial}
\nc{\na}{\nabla}
\nc{\ra}{\rangle}
\nc{\la}{\langle}
\nc{\goto}{\rightarrow}
\nc{\swap}{\leftrightarrow}

\nc{\EE}[1]{ \mbox{$\cdot10^{#1}$} }
\nc{\abs}[1]{\left|#1\right|}
\nc{\at}[2]{\left.#1\right|_{#2}}
\nc{\norm}[1]{\|#1\|}
\nc{\abscut}[2]{\Abs{#1}_{\scriptscriptstyle#2}}
\nc{\vek}[1]{{\rm\bf #1}}
\nc{\integral}[2]{\int\limits_{#1}^{#2}}
\nc{\inv}[1]{\frac{1}{#1}}
\nc{\dd}[2]{{{\partial #1}\over{\partial #2}}}
\nc{\ddd}[2]{{{{\partial}^2 #1}\over{\partial {#2}^2}}}
\nc{\dddd}[3]{{{{\partial}^2 #1}\over
        {\partial #2 \partial #3}}}
\nc{\dder}[2]{{{d #1}\over{d #2}}}
\nc{\ddder}[2]{{{d^2 #1}\over{d {#2}^2}}}
\nc{\dddder}[3]{{d^2 #1}\over
        {d #2 d #3}}
\nc{\dx}[1]{d\,^{#1}x}
\nc{\dy}[1]{d\,^{#1}y}
\nc{\dz}[1]{d\,^{#1}z}
\nc{\dl}[1]{\frac{d\,^{#1}l}{(2\pi)^{#1}}}
\nc{\dk}[1]{\frac{d\,^{#1}k}{(2\pi)^{#1}}}
\nc{\dq}[1]{\frac{d\,^{#1}q}{(2\pi)^{#1}}}

\nc{\cc}{\mbox{$c.c.$ }}
\nc{\hc}{\mbox{$h.c.$ }}
\nc{\cf}{cf.\ }
\nc{\erfc}{{\rm erfc}}
\nc{\Tr}{{\rm Tr\,}}
\nc{\tr}{{\rm tr\,}}
\nc{\pol}{{\rm pol}}
\nc{\sign}{{\rm sign}}
\nc{\bfT}{{\bf T }}

\def\GeV{{\rm\ GeV}}

\nc{\cA}{{\cal A}}
\nc{\cB}{{\cal B}}
\nc{\cD}{{\cal D}}
\nc{\cE}{{\cal E}}
\nc{\cG}{{\cal G}}
\nc{\cH}{{\cal H}}
\nc{\cL}{{\cal L}}
\nc{\cO}{{\cal O}}
\nc{\cT}{{\cal T}}
\nc{\cN}{{\cal N}}
\nc{\rvac}[1]{|{\cal O}#1\rangle}
\nc{\lvac}[1]{\langle{\cal O}#1|}
\nc{\rvacb}[1]{|{\cal O}_\beta #1\rangle}
\nc{\lvacb}[1]{\langle{\cal O}_\beta #1 |}
\nc{\bb}{\bar{\beta}}
\nc{\bt}{\tilde{\beta}}
\nc{\ctH}{\tilde{\cal H}}
\nc{\chH}{\hat{\cal H}}
\nc{\1}{\aa}
\nc{\2}{\"{a}}
\nc{\3}{\"{o}}
\nc{\4}{\AA}
\nc{\5}{\"{A}}
\nc{\6}{\"{O}}
\nc{\al}{\alpha}
\nc{\g}{\gamma}
\nc{\Del}{\Delta}
\nc{\e}{\epsilon}
\nc{\eps}{\epsilon}
\nc{\lam}{\lambda}
\nc{\om}{\omega}
\nc{\Om}{\Omega}
\nc{\ve}{\varepsilon}
\nc{\mn}{{\mu\nu}}
\nc{\k}{\kappa}
\nc{\vp}{\varphi}

\nc{\advp}[3]{{\it  Adv.\ in\ Phys.\ }{{\bf #1} {(#2)} {#3}}}
\nc{\annp}[3]{{\it  Ann.\ Phys.\ (N.Y.)\ }{{\bf #1} {(#2)} {#3}}}
\nc{\apl}[3]{{\it  Appl. Phys. Lett. }{{\bf #1} {(#2)} {#3}}}
\nc{\apj}[3]{{\it  Ap.\ J.\ }{{\bf #1} {(#2)} {#3}}}
\nc{\apjl}[3]{{\it  Ap.\ J.\ Lett.\ }{{\bf #1} {(#2)} {#3}}}
\nc{\app}[3]{{\it Astropart.\ Phys.\ }{{\bf #1} {(#2)} {#3}}}
\nc{\cmp}[3]{{\it  Comm.\ Math.\ Phys.\ }{{ \bf #1} {(#2)} {#3}}}
\nc{\cqg}[3]{{\it  Class.\ Quant.\ Grav.\ }{{\bf #1} {(#2)} {#3}}}
\nc{\epl}[3]{{\it  Europhys.\ Lett.\ }{{\bf #1} {(#2)} {#3}}}
\nc{\ijmp}[3]{{\it Int.\ J.\ Mod.\ Phys.\ }{{\bf #1} {(#2)} {#3}}}
\nc{\ijtp}[3]{{\it Int.\ J.\ Theor.\ Phys.\ }{{\bf #1} {(#2)} {#3}}}
\nc{\jmp}[3]{{\it  J.\ Math.\ Phys.\ }{{ \bf #1} {(#2)} {#3}}}
\nc{\jpa}[3]{{\it  J.\ Phys.\ A\ }{{\bf #1} {(#2)} {#3}}}
\nc{\jpc}[3]{{\it  J.\ Phys.\ C\ }{{\bf #1} {(#2)} {#3}}}
\nc{\jap}[3]{{\it J.\ Appl.\ Phys.\ }{{\bf #1} {(#2)} {#3}}}
\nc{\jpsj}[3]{{\it J.\ Phys.\ Soc.\ Japan\ }{{\bf #1} {(#2)} {#3}}}
\nc{\lmp}[3]{{\it Lett.\ Math.\ Phys.\ }{{\bf #1} {(#2)} {#3}}}
\nc{\mpl}[3]{{\it  Mod.\ Phys.\ Lett.\ }{{\bf #1} {(#2)} {#3}}}
\nc{\ncim}[3]{{\it  Nuov.\ Cim.\ }{{\bf #1} {(#2)} {#3}}}
\nc{\np}[3]{{\it  Nucl.\ Phys.\ }{{\bf #1} {(#2)} {#3}}}
\nc{\npps}[3]{{\it  Nucl.\ Phys.\ Proc.\ Suppl.\ }{{\bf #1} {(#2)} {#3}}}
\nc{\pr}[3]{{\it Phys.\ Rev.\ }{{\bf #1} {(#2)} {#3}}}
\nc{\pra}[3]{{\it  Phys.\ Rev.\ A\ }{{\bf #1} {(#2)} {#3}}}
\nc{\prb}[3]{{\it  Phys.\ Rev.\ B\ }{{{\bf #1} {(#2)} {#3}}}}
\nc{\prc}[3]{{\it  Phys.\ Rev.\ C\ }{{\bf #1} {(#2)} {#3}}}
\nc{\prd}[3]{{\it  Phys.\ Rev.\ D\ }{{\bf #1} {(#2)} {#3}}}
\nc{\prl}[3]{{\it Phys.\ Rev.\ Lett.\ }{{\bf #1} {(#2)} {#3}}}
\nc{\pl}[3]{{\it  Phys.\ Lett.\ }{{\bf #1} {(#2)} {#3}}}
\nc{\prep}[3]{{\it Phys.\ Rep.\ }{{\bf #1} {(#2)} {#3}}}
\nc{\prsl}[3]{{\it Proc.\ R.\ Soc.\ London\ }{{\bf #1} {(#2)} {#3}}}
\nc{\ptp}[3]{{\it  Prog.\ Theor.\ Phys.\ }{{\bf #1} {(#2)} {#3}}}
\nc{\ptps}[3]{{\it  Prog\ Theor.\ Phys.\ suppl.\ }{{\bf #1} {(#2)} {#3}}}
\nc{\physa}[3]{{\it  Physica\ A\ }{{\bf #1} {(#2)} {#3}}}
\nc{\physb}[3]{{\it  Physica\ B\ }{{\bf #1} {(#2)} {#3}}}
\nc{\phys}[3]{{\it Physica\ }{{\bf #1} {(#2)} {#3}}}
\nc{\rmp}[3]{{\it  Rev.\ Mod.\ Phys.\ }{{\bf #1} {(#2)} {#3}}}
\nc{\rpp}[3]{{\it Rep.\ Prog.\ Phys.\ }{{\bf #1} {(#2)} {#3}}}
\nc{\sjnp}[3]{{\it Sov.\ J.\ Nucl.\ Phys.\ }{{\bf #1} {(#2)} {#3}}}
\nc{\spjetp}[3]{{\it Sov.\ Phys.\ JETP\ }{{\bf #1} {(#2)} {#3}}}
\nc{\yf}[3]{{\it Yad.\ Fiz.\ }{{\bf #1} {(#2)} {#3}}}
\nc{\zetp}[3]{{\it Zh.\ Eksp.\ Teor.\ Fiz.\  }{{\bf #1}  {(#2)} {#3}}}
\nc{\zp}[3]{{\it Z.\ Phys.\ }{{\bf #1} {(#2)} {#3}}}
\nc{\ibid}[3]{{\sl ibid.\ }{{\bf #1} {#2} {#3}}}
\nc{\rf}[1]{(\ref{#1})}
\nc{\nn}{\nonumber \\*}
\nc{\bfB}{\bf{B}}
\nc{\bfv}{\bf{v}}
\nc{\bfx}{\bf{x}}
\nc{\bfy}{\bf{y}}
\nc{\vx}{\vec{x}}
\nc{\vy}{\vec{y}}
\nc{\oB}{\overline{B}}
\nc{\oI}{\overline{I}}
\nc{\oR}{\overline{R}}
\nc{\rar}{\rightarrow}
\nc{\ti}{\times}
\nc{\slsh}{\hskip-5pt/}
\nc{\sm}{Standard~Model~}
\nc{\MP}{M_{\rm Pl}}
\nc{\tp}{t_{\rm Pl}}
\nc{\ave}{\bar{E}}

\nc{\eff}{{\rm eff}}
\nc{\kk}{\vek{k}}
\nc{\pp}{{\rm p}}
\nc{\ga}{g_{a\gamma}}
\nc{\vv}{\\}
\nc{\eee}{{\bf E}}
\nc{\bbb}{{\bf B}}
\nc{\qcd}{T_{\rm QCD}}
\nc{\G}{\rm \ G}
\def\vec#1{{\bf #1}}

\def\lae{\;^{<}_{\sim} \;}  

\def\ell{e^{c}LL}

\begin{document}
{\title{\vskip-2truecm{\hfill {{\small \\
        \hfill \\
        }}\vskip 1truecm}
{\LARGE  F-term Hybrid Inflation, the
$\eta$-Problem and Extra Dimensions}}
{\author{
{\sc  John McDonald$^{1}$}\\
{\sl\small Theoretical Physics Division,
University of Liverpool,
Liverpool L69 3BX, England}
}
\maketitle
\begin{abstract}
\noindent

           F-term hybrid inflation models in the context of supergravity
generically have corrections to the inflaton mass squared of the order of 
$H^{2}$ (the $\eta$-problem). 
In addition they have a problem with large deviations of the 
spectrum of density perturbations from scale-invariance 
due to non-renormalizable corrections to the superpotential. 
Here we show that an increase of the expansion
 rate at large energy densities relative to that in conventional
 $d=4$ cosmology, as suggested by 
brane scenarios, can naturally solve the $\eta$ problem.
For the case of a $\rho^{2}$ squared correction to $H^{2}$ suppressed by the brane tension 
this requires that 
the $d=5$ Planck mass satisfies $M_{5} \lae 10^{16} \GeV$. In addition,
 the scale-invariance of the density perturbation spectrum 
can be generally protected from Planck-suppressed 
superpotential corrections if 
$M_{5} \lae 10^{10} \GeV$. Therefore brane cosmologies which
produce an enhanced expansion rate at large energy densities are
favoured by SUSY hybrid inflation.

\end{abstract}
\vfil
\footnoterule
{\small $^1$mcdonald@sune.amtp.liv.ac.uk}

\thispagestyle{empty}
\newpage
\setcounter{page}{1}

\section{Introduction}

          Hybrid inflation  models \cite{hi} 
are a favoured class of inflation model, being able to 
account for both the flatness 
of the inflaton potential and the rapid inflaton
 oscillations which end inflation, without the need for very small couplings.
 Supersymmetric (SUSY) versions 
are classified as F- \cite{fti,cope} or D-term \cite{dti}
 hybrid inflation models, depending on the origin of the 
energy density driving inflation. However, once globally 
 SUSY models are generalized to supergravity 
(SUGRA), a number of problems arise. For the case of F-term hybrid inflation
 models, 
there are corrections to the inflaton mass squared of the order
 of the expansion rate squared arising from SUGRA
 corrections \cite{eta,cope}. Although it is possible 
to eliminate such corrections by choosing the
 K\"ahler potential carefully, such as 
having a minimal form \cite{cope} or a Heisenberg
 symmetry \cite{oliveh}, in most 
models such corrections result in an inflaton potential 
which cannot produce successful slow-roll
 inflation (the $\eta$-problem).  
In addition, SUSY hybrid inflation models
 (both F- and D-term) have a problem with 
deviations of the inflaton potential from flatness
 due to Planck-suppressed non-renormalizable
 corrections to the superpotential. 
This arises due to the large initial inflaton expectation
 value, typically close to the Planck scale, required
 in order to achieve sufficient inflation \cite{super,lr}. 

         All this assumes the standard $d=4$ cosmology
 based on the homogeneous and isotropic
 Friedmann-Robertson-Walker (FRW) metric.
 In the past few years there has been considerable interest 
in the possibility that spacetime may have more than
 four dimensions \cite{exd,rs1,rs2}, which is 
possible if conventional matter is confined to a $d=4$ brane. 
For example, in the Randall-Sundrum (RS) brane scenario with a single brane and one 
uncompactified extra dimension (RS2) \cite{rs2}, the brane
 has a positive tension and the 
$d=5$ bulk a negative cosmological constant such that the sum cancels
 out on the brane, giving zero (or very small, if the cancellation
 is not exact) cosmological constant 
in the $d=4$ Universe. Conventional gravity is 
obtained on the brane by localizing the massless graviton mode via 
the near-brane geometry.
The cosmology of this scenario is no
longer described by the $d=4$ FRW metric but instead
 one must solve the $d=5$ Einstein equations for the
 evolution of the full $d=5$ metric. 
 The solution of the Einstein equations in the RS2 scenario  
including matter trapped on the brane results in
 a different evolution for the $d=4$ expansion rate 
$H$ as a function of the energy density $\rho$ as compared
 with conventional $d=4$ cosmology \cite{bine,flan}, such
 that at large energy density $H$ is proportional to
 $\rho$ rather than $\rho^{1/2}$. It follows that
 inflationary cosmology has to be reconsidered in the RS2 brane scenario. 
This has recently been done for the case of chaotic
 inflation, where it was found that the brane
 has the advantage of reducing the
 magnitude of the inflaton field during inflation
 to below the Planck scale, so evading dangerous
 Planck-suppressed non-renormalizable corrections
 to the inflaton potential \cite{cbrane,bb}.

               However, it has been shown that the single-brane
 RS2 scenario is not
 compatible with SUSY \cite{rssusy}. Nevertheless, we generally 
expect to find corrections to the expansion rate as a function of
the energy density in brane scenarios \cite{cline,fr1}\footnote{
It has also been noted that, due to
higher-derivative corrections to the gravitational action, 
the $\rho^{2}$ correction to $H^{2}$
must probably be considered as 
the first term in an expansion in energy density 
suppressed by powers of the brane tension, 
in which case we cannot be sure of the form of $H^{2}$ at 
large energy densities in the absence of a 
full quantum theory of gravity (even in the case of the RS2 scenario) 
\cite{cline}.}. Generalized brane
corrections of the form $H^{2} = A \rho + B \rho^{n}$
have recently been suggested in order to account for the 
observed accelerated expansion of the Universe \cite{fr2}.
Any relation between $H^{2}$ and $\rho$ is in principle
possible in brane cosmology, depending on the bulk
stress-energy tensor and boundary conditions \cite{fr1}, the simplest case being 
that of a cosmological constant in the bulk and 
a cancelling positive tension on the brane, which results in a $\rho^{2}$ correction 
suppressed by the brane tension \cite{cline}.
In order to explore the possible advantages for
SUSY inflation of an
 enhanced expansion rate at large energy 
densities, in the following we will consider
 the case of $N=1$ $d=4$ SUGRA inflation but with a $\rho^{2}$ correction 
to the expansion rate suppressed by the brane tension, using the RS2 correction as a 
specific example. 
We will refer to this as the "brane-inspired 
scenario". (It should again be emphasized that the RS2 model does not 
apply in the SUSY case; we use the Friedmann equation of the RS2 model 
only as an example of what might arise in brane cosmology.) 
In particular we will consider the case of F-term hybrid inflation. 
We will show that the $\eta$ 
problem can be easily and naturally solved
 in this scenario and, in addition, that the 
problem of corrections to the inflaton superpotential
 may be solved if the fundamental 
$d=5$ Planck mass $M_{5}$ is sufficiently small.

         The letter is organized as follows. In Section 2
 we discuss the F-term hybrid inflation model in the brane-inspired scenario. 
In Section 3 we review the $\eta$-problem
 and its solution in the brane-inspired scenario. 
In Section 4 we consider the problem of
 non-renormalizable corrections to 
the superpotential and its solution in the brane-inspired scenario.
 In Section 5 we present our conclusions. 

\section{F-term Hybrid Inflation in the Brane-Inspired Scenario} 

        The F-term hybrid inflation model is described
 by the superpotential \cite{fti}
 \be{e2} W = S\left(\kappa \overline{\psi}\psi - \mu^{2}\right)       ~.\ee
The tree-level scalar potential in the global SUSY limit is then 
\be{e3} V = |\kappa \overline{\psi}\psi - \mu^{2}|^{2} + 
\kappa^{2} |S|^{2}\left( |\overline{\psi}|^{2} + |\psi|^{2}\right)      ~.\ee
We can take the inflaton field to be
 real and positive. If $s > s_{c} \equiv \sqrt{2}\mu/\kappa^{1/2}$
 (where $<S> = s/\sqrt{2}$) then at the minimum of 
$V$ as a function of $s$ we find $\psi = \overline{\psi} = 0$
 and so the tree-level potential as a function of $s$ is flat, 
$V(s) = \mu^{4}$. Including the one-loop correction due
 to $\psi$ and $\overline{\psi}$
loops results in the potential (for $s^{2}$ large compared
 with $s_{c}^{2}$) \cite{fti} 
\be{e4} V(s) \approx \mu^{4}\left( 1
 + \frac{\kappa^{2}}{16 \pi^{2}} ln \left(\frac{s}{Q}\right)  
\right)    ~,\ee
where $Q$ is a renormalization scale.
The slow-roll equation for $s$ is then 
\be{e5} 3H \dot{s} = - \frac{\partial V}{\partial s} =
- \frac{\kappa^{2} \mu^{4}}{16 \pi^{2} s}    ~.\ee
The Friedmann equation for the $d=4$ expansion rate
 $H$ in the RS2 brane scenario is given by \cite{bine,flan}
\be{e1} H^{2} = \frac{\Lambda_{4}}{3} 
+ \left(\frac{8 \pi}{3 M_{4}^{2}}\right) \rho
+ \left( \frac{4 \pi}{3 M_{5}^{3}} \right)^{2} \rho^{2}
+ \frac{\epsilon}{a^{4}}    ~,\ee  
where $\Lambda_{4}$ is the $d=4$ cosmological 
constant, $\epsilon$
is due to the influence of graviton modes 
in the bulk and $M_{4}$ and $M_{5}$ 
are the $d=4$ and $d=5$ Planck masses respectively. 
For the RS2 brane scenario $M_{5}$ must be 
greater than $10^{8} \GeV$ in order that gravity 
is not modified on length scales greater than 
the experimental limit of about 
1mm \cite{cbrane}. (We will see that $M_{5}$ is
generally larger than this in following.)
The $\epsilon$ term is
diminished during inflation by expansion and may
 be neglected, as may the small $d=4$ cosmological
 constant. Therefore during inflation in the RS2 scenario
\be{e6} H^{2} = \left( \frac{4 \pi}{3 M_{5}^{3}} \right)^{2}
 \rho \left(\rho + \rho_{c} \right) \equiv  
 \frac{8 \pi \rho}{3 M_{4}^{2}} \left(1 + \frac{\rho}{2 \tau}\right) ~,\ee  
where $\tau = 48 \pi M_{5}^{6}/M_{4}^{2}$ is the 3-brane tension and 
\be{e7} \rho_{c} = \frac{3}{2 \pi} \frac{M_{5}^{6}}{M_{4}^{2}}      ~.\ee
$H^{2}$ is dominated by the $\rho^{2}$ term if
 $\rho \approx \mu^{4} > \rho_{c}$.
We will assume this form for $H^{2}$ in the following as a specific example 
of a brane-generated $\rho^{2}$ correction to $H^{2}$ suppressed by the 3-brane tension. 

      Using $H^{2}$ in \eq{e6}, the solution to the
slow-roll equation \eq{e5} is given by 
\be{e8} s^{2} - s_{c}^{2}  
 = \frac{N \kappa^{2} }{24 \pi^{2} \left(\rho + \rho_{c}\right) } 
\left( \frac{3 M_{5}^{3}}{4 \pi} \right)^{2}   ~,\ee 
where $N$ is the number of e-foldings of inflation as $s$
 rolls from $s$ to $s_{c}$
(assuming $H$ is approximately constant during inflation). 
The adiabatic density perturbations evolve as
 in conventional $d=4$ cosmology \cite{cbrane,dp}, 
with $\zeta \propto \delta \rho/\left(\rho + p \right)$
 conserved on superhorizon scales, 
such that $\delta \rho/\rho$ for a density
 perturbation mode re-entering the horizon is
 proportional to $\delta \rho/\dot{\phi}^{2}$
 on horizon crossing during inflation. 
During slow-rolling the COBE normalization
 of the density perturbation is given by \cite{cobe}
\be{e9} \delta_{H} \equiv \frac{3}{5 \pi} \frac{H^{3}}{V^{'}}
= \delta_{H \; COBE} \equiv 1.91 \times 10^{-5}   ~,\ee
where $\delta_{H}$ is the density perturbation
 spectrum at the present Hubble scale \cite{cobe}. 
COBE normalization then implies that 
\be{e10} \mu = \left(\frac{15}{256 \pi^{2}}\right)^{1/6}
 \left(\frac{24}{N}\right)^{1/12} 
\kappa^{1/6} \delta_{H \; COBE}^{1/6}
 \left(1 + \frac{\rho_{c}}{\rho} \right)^{-1/6}
  M_{5}  ~.\ee 
Thus 
\be{e11} \mu = 0.065  \kappa^{1/6} 
\left( \frac{50}{N} \right)^{1/12} 
\left(1 + \frac{\rho_{c}}{\rho} \right)^{-1/6}M_{5}     ~,\ee
where $N \sim 50$ is typical of the number of e-foldings
 at which COBE length scales exit the horizon.  

     In addition to imposing COBE normalization
 we will assume that $s^{2}$ is large compared with
 $s_{c}^{2}$ during inflation, in order that \eq{e4} is valid. 
With $s_{o}$ denoting the inflaton field when the observable Universe 
leaves the horizon ($N \sim 50-60$), this requires that 
\be{e12} s_{o}^{2} \approx \frac{N \kappa^{2}}{24 
\left(\rho + \rho_{c}\right) \pi^{2}} 
\left( \frac{3 M_{5}^{3}}{4 \pi} \right)^{2} \gg s_{c}^{2}   ~.\ee
This is satisfied if 
\be{e13} \mu < \left( \frac{N \kappa^{3}}{48 \pi^{2}} \right)^{1/6} 
\left( \frac{3 M_{5}^{3}}{4 \pi} \right)^{1/3}
 \left( 1 + \frac{\rho_{c}}{\rho} \right)^{-1/6}    ~,\ee
with $s_{c}^{2}/s_{o}^{2} \propto \mu^{6}$.  

      Finally, in order to have a scenario that deviates from
 conventional $d=4$ F-term hybrid inflation we have
 to impose that $\rho > \gamma \rho_{c}$ during
 inflation ($\gamma$ is a constant which will be
 useful later; for now we may take $\gamma = 1$), which requires that 
\be{e14} M_{5} < \left(\frac{2 \pi 
\rho M_{4}^{2}}{3 \gamma}\right)^{1/6}    ~.\ee

       We can combine these conditions to obtain
 constraints on the coupling $\kappa$ and the $d=5$
 Planck mass $M_{5}$. The COBE
 normalization, \eq{e11}, combined with the
 condition for $s_{o}^{2} \gg s_{c}^{2}$, \eq{e13}, 
implies the $M_{5}$ independent condition 
\be{e16} \kappa > 3.5 \times 10^{-3} 
\left(\frac{50}{N}\right)^{3/4}    ~.\ee
The condition $\rho > \gamma \rho_{c}$, \eq{e14}, combined
 with the COBE normalization, \eq{e11}, implies that 
\be{e17} \gamma^{1/2} M_{5} < 7.4 \times 10^{16}
 \kappa^{1/3}  \left(\frac{50}{N}\right)^{1/6} \GeV  \equiv M_{5\;c}  ~,\ee
(where for $M_{5} \geq M_{5\;c}$ we have conventional
 $d=4$ cosmology during inflation).
Thus from \eq{e16} and \eq{e17}, with $\gamma = 1$
 and $\kappa$ of the order of 1, we see that
 successful F-term hybrid inflaton with the 
expansion rate dominated by brane effects can be naturally achieved if 
$M_{5} \lae 10^{16-17} \GeV$.  

\section{$\eta$-Problem in the Brane-Inspired Scenario}

          In SUGRA models the F-term contribution to the
 scalar potential is given by
\cite{lr} 
\be{19} V = F^{2} - \frac{3}{M^{2}}e^{K/M^{2}}|W|^{2}    ~,\ee
where $W$ is the superpotential, $K$ is the
 K\"ahler potential, $M = M_{4}/\sqrt{8\pi}$ is the SUGRA mass scale and 
\be{e20} F^{2} = e^{K/M^{2}} \left(W_{m}
 + \frac{W K_{m}}{M^{2}}\right)^{*}
K^{m^{*}\;n}\left(W_{n} + \frac{W K_{n}}{M^{2}}\right)    ~,\ee
where $W_{m} = \partial W/\partial \phi^{m}$ and
 $K^{m^{*}n}$ is the inverse of $K_{nm^{*}}$.
For example, with the superpotential given by \eq{e2}
 and with K\"ahler potential 
for the inflaton $K = |S|^{2} + \Delta K$, 
where $\Delta K = c|S|^{4}/M^{2}$ and $c$ is a dimensionless
 constant and so expected to be of the order of 1, 
the SUGRA scalar potential for $s > s_{c}$ is given by 
\be{e21} V = V_{susy} - \frac{2 c \mu^{4}s^{2}}{M^{2}}  
+ \left(\frac{\mu^{4} s^{4}}{M^{4}}\right)  ~,\ee
where $V_{susy}$ is the global SUSY scalar
 potential. Thus except 
in the case of $c$ equal to 0 
(corresponding to a minimal K\"ahler potential), 
the inflaton typically gains a mass squared 
correction $\sim \pm \mu^{4}/M^{2}$. 
In the case of conventional 
$d=4$ cosmology, the expansion rate during inflation is 
$H^{2} = \mu^{4}/3 M^{2}$. Therefore the
 magnitude of the mass squared correction is of the order of $H^{2}$. 
However, it should be emphasized this is {\it essentially coincidental}. 
For the case of the brane-inspired scenario, the mass squared correction 
remains $\sim \pm \mu^{4}/M^{2}$ but, 
for energy density $\rho > \rho_{c}$, $H^{2}$
 from \eq{e6} is now larger than $\mu^{4}/3M^{2}$
 by a factor $\rho/\rho_{c}$. 
Thus so long as $\rho/\rho_{c}$ is sufficiently
 large (greater than about 300), 
the mass squared correction to the inflaton potential
 will be less than $0.1 H$ and so 
will not play a significant role in the dynamics
 of the inflaton. (The magnitude of $\eta$-parameter (defined in \eq{e24} below)
due to the mass squared correction is then $\lae 1/300$ and so
completely negligible.)
From \eq{e17} with $\gamma = 300$ we see that 
this requires that $M_{5} \lae 10^{16} \GeV$. Therefore
so long as the scale of gravity in 
the bulk is sufficiently small, the $\eta$-problem
 of F-term hybrid inflation models is solved 
in the brane-inspired scenario. 

\section{Non-renormalizable Superpotential Corrections} 

             In conventional $d=4$ cosmology 
there is a second problem for SUSY hybrid inflation models 
associated with the large initial value of the 
inflaton field required to account for 
sufficient inflation \cite{super}. From \eq{e8}, 
if $\rho < \rho_{c}$, corresponding to conventional $d=4$
 cosmology during inflation, then the value of $s_{o}$
 (for $s_{o}^{2}$ large compared with $s_{c}^{2}$), is given by 
\be{e22} s_{o} = \frac{\kappa N^{1/2}M}{2 \sqrt{2} \pi}     ~.\ee
Thus in conventional $d=4$ cosmology and with $\kappa$
 of the order of 1, the initial 
expectation value of the inflaton is of the order of $M$. 
Since in SUGRA there is no reason to restrict 
the superpotential to renormalizable terms, 
we expect in general to find Planck-suppressed
 non-renormalizable superpotential terms of the form 
$\lambda_{n} S^{n}/n!M^{n-3}$ (where we
 include a symmetry factor $n!$). This results
 in a correction to the inflaton potential 
\be{e23}  \delta V=  \frac{ \lambda_{n}^{2} 
|S|^{2(n-1)}}{((n-1)!)^{2} M^{2(n-3)}}    ~.\ee
Since in $d=4$ cosmology $s$ is typically not very
 small compared with $M$, this results in a large
 deviation of the spectrum of density perturbation
 from scale-invariance \cite{super}.
The spectral index is given by $n = 1 + 2 \eta - 6 \epsilon$, where
in general during slow-rolling
\be{e24} \eta = \frac{1}{3} \frac{V^{''}}{H^{2}} \;,\;\;\;\; 
\epsilon = \frac{1}{3} \frac{V^{' \; 2}}{H^{3}}
  \frac{\partial H}{\partial V}    ~,\ee    
with prime denoting $\partial/\partial s$. (Observationally
 $|n-1| \lae 0.1 $ \cite{obs}.) For the 1-loop 
correction, \eq{e4}, $\eta = -1/2N$ and 
$|\epsilon/\eta| \approx \kappa^{2}/16\pi^{2} \ll 1$. Assuming that the
 contribution of the superpotential correction $\delta V$ to $V^{''}$ 
dominates that of the 1-loop correction, 
$\eta$ is then given by  
\be{e25} \eta = \frac{2}{3} \left( \frac{3 M_{5}^{3}}{4 \pi}\right)^{2} 
\frac{(n-1)(2n-3) 
\tilde{\lambda}_{n}^{2}s^{2n-4}}{\rho \left(\rho 
+ \rho_{c}\right)M^{2(n-3)}}     ~,\ee
where $\tilde{\lambda}_{n} = \lambda_{n}/(2^{(n-1)/2} (n-1)!)$.  
(For $V^{'}$, $V^{''}$ dominated by $\delta V$, $|\epsilon/\eta| \approx 
\delta V/V < 1$. Therefore $\eta$ 
gives the dominant contribution to the index.)  
Imposing the constraint that $  |2 \eta| \lae |\Delta n| $ then implies 
\be{e26} \left(\frac{s}{M}\right)^{2n-4}  
< \frac{3}{4} \left(\frac{4 \pi}{3 M_{5}^{3}}
\right)^{2} \frac{ \rho(\rho + \rho_{c}) \Delta n}{(n-1)(2n-3) 
\tilde{\lambda}_{n}^{2} M^{2}}
~.\ee
The largest non-renormalizable correction corresponds to $n=4$. 
From \eq{e12}, the value of $s_{o}$ for 
$\rho > \rho_{c}$ and $s_{o}^{2} \gg s_{c}^{2}$ is 
\be{e27} s_{o}^{2} = \frac{N \kappa^{2}}{24 \pi^{2}\rho} 
 \left(\frac{3 M_{5}^{3}}{4 \pi}\right)^{2}    ~.\ee
Thus with $s = s_{o}$ in \eq{e26} we 
obtain for $n = 4$ an upper limit on $M_{5}$,  
\be{e28} M_{5} < 1.9 \times 
10^{10}\left(\frac{1}{\tilde{\lambda}_{4} \kappa^{2/3}}
\right) \left(\frac{50}{N}\right)^{5/3} 
\left(\frac{\Delta n}{0.1} \right)^{1/2}  
\GeV    ~.\ee 
Thus for all $\kappa \lae 1$, if  $M_{5} \lae 10^{10-11} \GeV$ then  
{\it all} non-renormalizable corrections are compatible with
 scale-invariance of the spectrum of density
 perturbations. Note that from COBE normalization, \eq{e11}, the
 corresponding scale of the energy density 
driving inflation, $V^{1/4} = \mu$, is then
$\mu \lae 10^{9-10} \GeV$. 
Therefore the thermal gravitino upper bound
 on the reheating temperature, $T_{R} \lae 10^{8}
 \GeV$ \cite{grav}, should be easier to satisfy in this case. 

     In general, even if Planck-suppressed non-renormalizable superpotential 
corrections are acceptable in the brane-inspired scenario, renormalizable 
$S^{2}$ and $S^{3}$ terms must still be eliminated. 
Although it is possible to eliminate all superpotential corrections 
(both renormalizable and non-renormalizable) via a global R-symmetry, 
it has been argued that global symmetries are generally broken by
 wormhole \cite{worm} or string \cite{string} effects,
 leaving only discrete symmetries as a possibility
 for eliminating unwanted superpotential terms. 
In this case it is possible that at least some non-renormalizable superpotential
 corrections will be allowed by the discrete symmetry. 
 Therefore the ability to tolerate unsuppressed 
non-renormalizable corrections may be a significant 
advantage for enhanced expansion at large energy densities. 

\section{Conclusions} 

     We have discussed how F-term hybrid
 inflation in the context of $N=1$ supergravity models 
may be modified advantageously via an enhanced
 expansion rate as a function of $\rho$, such as may occur in brane cosmology.
 As a specific model of a $\rho^{2}$ 
correction to $H^{2}$ suppressed by powers
 of the 3-brane tension 
 we assumed the form of the Friedmann equation
 obtained in the (non-SUSY) RS2 scenario. (This is used here only to 
model the form of expansion rate that might plausibly be obtained in brane
 cosmology and in order to illustrate the advantages
 of an enhanced expansion rate for F-term inflation.) 
 If the $d=5$ Planck scale 
$M_{5}$ is less than about $10^{16} \GeV$ 
then SUGRA corrections to the inflaton
mass are less than 0.1$H$, so solving the $\eta$ problem. 
In addition, if $M_{5} \lae 10^{10-11} \GeV$, 
then Planck-suppressed non-renormalizable 
superpotential corrections do not cause
 an unacceptable deviation of the density perturbation
 spectrum  from scale-invariance. 
In this case the energy density during inflation
 satisfies $\rho^{1/4} \lae 10^{9-10}\GeV$,
 making it easier for the reheating temperature
 to satisfy the thermal gravitino upper bound. Although we have 
discussed a specific form for the enhanced expansion rate, 
 we expect that the benefits for SUSY hybrid inflation will be obtained 
for enhanced expansion rates in general. 
Thus an increased expansion rate as
compared with conventional $d=4$ cosmology, such as may occur in brane 
cosmology, is favoured by SUSY F-term hybrid inflation. This may serve
 as a motivation for a finding a SUSY brane cosmology with a suitably enhanced
 expansion rate at large energy densities.

\end{document}